\documentclass[doc]{article}

\usepackage{amsmath,amssymb,amsthm}
\usepackage{graphicx}
\usepackage{float}
\usepackage{booktabs}
\usepackage{geometry}
\usepackage{cite}
\usepackage{caption}
\usepackage{subcaption}
\usepackage{url}
\usepackage[hidelinks]{hyperref}
\usepackage{balance}

\newenvironment{paperabstract}{%
  \vspace{6pt}%
  \noindent\rule{\linewidth}{0.4pt}\par\vspace{4pt}%
  \noindent\textbf{Abstract.}\enspace\ignorespaces
}{%
  \par\vspace{4pt}%
  \noindent\rule{\linewidth}{0.4pt}\par\vspace{10pt}%
}

\title{Geometric Analysis of the Damped Harmonic Oscillator\\
via the Lambert $W$ Function}

\author{
  Arpan Sharma$^{1}$ \quad Bhargava R. Jogi$^{2}$ \quad
  Ken Roberts$^{3}$  \quad
  Muralikrishna Molli$^{1}$ \quad S.R.\ Valluri$^{3,4}$\\[4pt]
  \small $^{1}$Department of Physics, Sri Sathya Sai Institute of Higher Learning,\\
  \small Prasanthi Nilayam, India\\[2pt]
  \small $^{2}$Wake Forest University, Winston-Salem, NC, USA\\[2pt]
  \small $^{3}$Department of Physics and Astronomy,\\
  \small University of Western Ontario (UWO), London, ON, Canada\\[2pt]
  \small $^{4}$Department of Mathematics, University of Western Ontario (UWO),\\
  \small and MEM, King's University College, UWO, London, ON, Canada\\[4pt]
  \small Correspondence: \texttt{krobe8@gmail.com , jogib24@wfu.edu }
}
\date{}   

\begin{document}

\maketitle
\begin{paperabstract}
The underdamped harmonic oscillator is analyzed through the complex
mapping $\zeta = e^{-i\varphi}we^{-w}$ with $w = \beta t + i\Omega t$,
which transforms the dynamics into a logarithmic spiral.
Within this framework, the displacement extrema correspond to crossings
of the imaginary axis by $\zeta(t)$, yielding the explicit times
$t_n = (\theta - \varphi - \pi/2 + n\pi)/\Omega$, where
$\theta = \arctan(\Omega/\beta)$.
The Lambert $W$ function provides closed-form solutions
$t = -\beta^{-1}W_k(-\beta A/\omega_0)$ for the times at which the
spiral radius attains a given threshold $A$, covering both the rising
and decaying branches.
The quality factor $Q = \omega_0/(2\beta) = \tfrac{1}{2}\sec\theta$ is
directly encoded in the ray angle $\theta$ of the $(u,v)$-plane.
Key geometric invariants are derived: the winding number
$N_\varepsilon \approx (Q/\pi)\ln(2Q/\varepsilon)$ for large $Q$,
the enclosed area $A = \omega_0^2\Omega/(8\beta^3) \approx Q^3$ in the
lightly damped limit, and the energy decay $E(t) = E_0 e^{-\omega_0 t/Q}$.
Three methods for determining $Q$ from experimental data are compared:
logarithmic decrement, ray-angle measurement, and spiral turn counting.
The turn-counting method proves particularly robust for high-$Q$ systems,
where successive amplitude peaks differ by tiny fractions.
The framework unifies classical damped oscillations with complex analysis
and special functions.
\end{paperabstract}

\section{Introduction}

The damped harmonic oscillator ranks among the most pervasive dynamical
systems in physics and engineering. Its governing equation---a
second-order linear ODE with constant coefficients---appears wherever
inertia, a restoring force, and dissipation coexist, underpinning the
design of everything from earthquake-resistant structures to
quantum-limited amplifiers.

\subsection*{Mechanical and Structural Applications}
In mechanical engineering the oscillator models vehicle
suspensions~\cite{thomson1998}, vibration isolators for sensitive
instruments, and the free-decay response of civil structures after
impulsive loading. The quality factor $Q$ is inversely proportional to
the fraction of vibrational energy dissipated per cycle, so optimising
$Q$ is synonymous with controlling fatigue, noise, and resonance
amplification.

\subsection*{Electrical Engineering: RLC Circuits}
The mechanical analogy is exact: inductance $L$ plays the role of
mass $M$, capacitance $C_{\mathrm{e}}$ the role of compliance $1/K$,
and resistance $R_{\mathrm{e}}$ the role of the damping constant $C$.
The quality factor governs bandwidth, selectivity, and the rate at
which stored electromagnetic energy is dissipated as heat. High-$Q$
RLC filters are the backbone of radio receivers, impedance-matching
networks~\cite{pozar2011}, and analogue signal processing chains.
\subsection*{Microwave, RF Cavities, and Particle Accelerators}
Superconducting radiofrequency (SRF) cavities in modern accelerators
such as CERN's Large Hadron Collider sustain intrinsic quality factors
of $Q_0 \sim 10^{10}$--$10^{11}$~\cite{padamsee2008}. Even a marginal
reduction in $Q_0$ translates directly into increased cryogenic heat
load and reduced accelerating gradient. In satellite and terrestrial
telecommunications, coupled-cavity waveguide filters achieve $Q$ values
of $10^3$--$10^5$, with passbands and group-delay responses determined
entirely by the $Q$ factors of the constituent resonators. The
Lambert~$W$ function has recently appeared in the analysis of
coupled-resonator chains~\cite{jenn2005}.

\subsection*{Quantum Computing: Superconducting Qubits}
The superconducting transmon qubit~\cite{koch2007} is implemented as a
Josephson-junction-shunted LC oscillator whose energy relaxation time
$T_1$ is Purcell-limited by the photon lifetime of its readout
resonator, bounding the coherence time as $T_2 \leq 2T_1$. Purcell
decay is suppressed by engineering the resonator $Q$~\cite{reed2010}
to lie in a precise window: high enough to protect the qubit from
radiative decay, yet low enough to permit fast, high-fidelity
readout~\cite{houck2008}.

\subsection*{Cavity Optomechanics}
In cavity optomechanical systems~\cite{aspelmeyer2014} a mechanical
oscillator is coupled to an optical or microwave cavity mode, and both
sub-systems obey the same damped oscillator equation. Their $Q$ factors
simultaneously govern the resolved-sideband condition
$\Omega \gg \kappa$, where $\kappa = \omega_0/Q$ is the cavity
linewidth, which must be satisfied to reach the quantum ground state
via laser cooling.
\subsection{The Lambert \texorpdfstring{$W$}{W}
Function and This Work}
The Lambert $W$ function\cite{mezo2022}, defined by $W(z)e^{W(z)}=z$ and
studied systematically by Corless et al.~\cite{corless1996,scott2006},
has found applications in delay-differential equations and
control systems~\cite{asl2003}, combinatorics, and
physics~\cite{valluri2000}. To our knowledge, a systematic
geometric development linking the branches of $W$ to
physically distinct instants in the oscillator's evolution
has not previously been given.

In this paper we develop such a framework, centred
on the complex spiral
\begin{equation}
    \zeta(t) = e^{-i\varphi}\,w\,e^{-w},
    \qquad w = \beta t + i\Omega t,
\end{equation}
and establish four principal results:
\begin{enumerate}
  \item Displacement extrema correspond precisely to
    imaginary-axis crossings of the spiral $\zeta(t)$,
    expressible in closed form via the two real branches
    of the Lambert $W$ function: $W_0$ gives the
    rise time and $W_{-1}$ gives the
     decay time (Section~4).

  \item The quality factor is encoded in the spiral in
    two equivalent geometric forms:
    $Q = \tfrac{1}{2}\sec\theta$, where $\theta$ is the
    angle of the evolution ray in the $(u,v)$-plane, and
    $Q = \tfrac{1}{2}\sqrt{1+(2\pi N_{\mathrm{turns}})^2}$,
    where $N_{\mathrm{turns}}$ is the number of spiral
    windings per amplitude decay time --- so that the
    ray angle and the winding count are two faces of the
    same geometric quantity (Sections~3 and~4).
    
    \item The spiral's winding number \(N_\varepsilon \approx (Q/\pi)\ln(2Q/\varepsilon)\) for large \(Q\) provides a robust method for estimating \(Q\) from experimental data by simply counting visible oscillations above a noise threshold. Unlike the logarithmic decrement, which requires precise amplitude ratios of successive peaks, turn counting remains reliable even when successive amplitudes differ by only a tiny fraction
  \item Mechanical energy decays as $E(R) = E_0 e^{-R/Q}$,
    where $R = \omega_0 t$ is the radial coordinate in the
    $(u,v)$-plane, recovering the standard definition of
    $Q$ as stored energy divided by energy lost per radian.

\end{enumerate}

\subsection{Paper Organisation}
The remainder of this paper is organised as follows.
Section~2 reviews the spring--mass--damper system
and its three damping regimes. Section~3 introduces
the Lambert~$W$ connection and the $(u,v)$-plane
geometry. Section~4 develops the $\zeta$-plane
spiral and its properties. Section~5 illustrates the
spiral across all damping regimes.
Section~6 analyses energy decay and its geometric
interpretation on the $(u,v)$-plane. Section~7
concludes with directions for extension to driven
oscillators and coupled RLC circuits.

\section{Spring--Mass System}
\label{sec:intro}
\subsection{Setup}
\begin{figure}[H]
  \centering
  \includegraphics[width=0.72\linewidth]{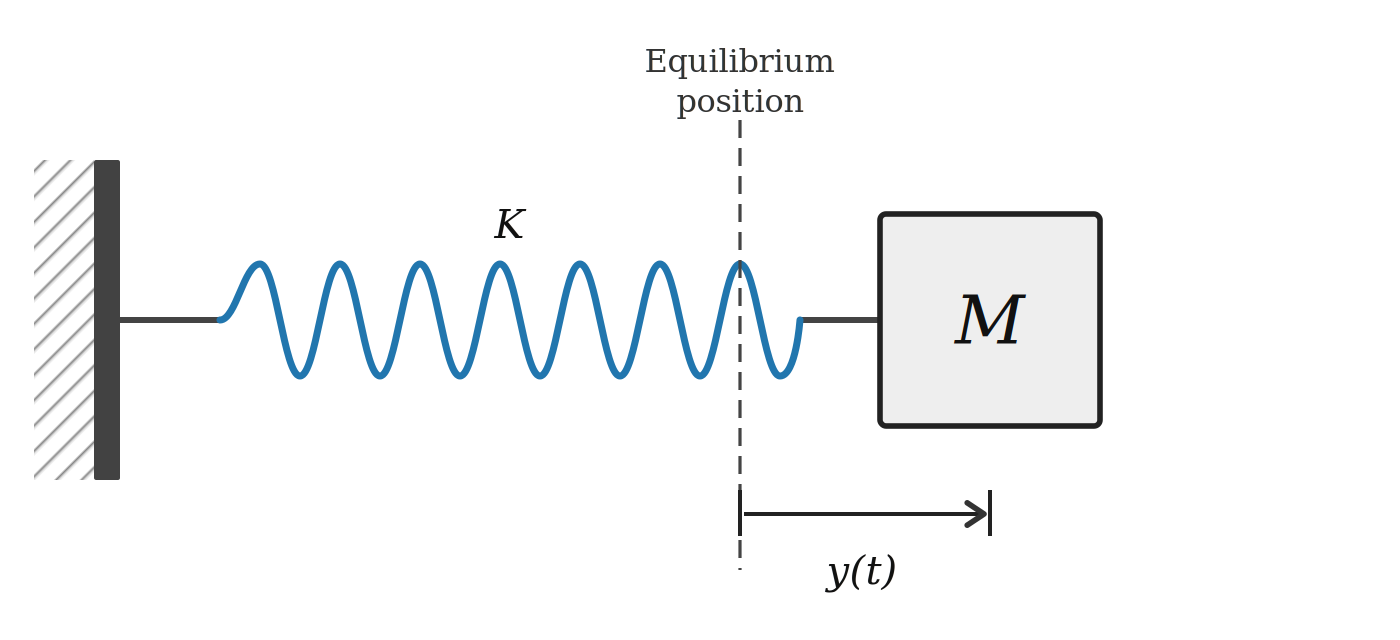}
  \caption{Spring--mass--damper system.}
  \label{fig:smd}
\end{figure}

We consider a mass--spring system in a viscous medium (Fig.~\ref{fig:smd}),
where $K$ is the spring constant and $C$ is the damping constant.
The viscous damping force is proportional to the velocity $\dot{y}$,
and the restoring force in the spring is $Ky$, where $y$ is the
displacement from the unstretched position.

Applying Newton's second law yields the second-order homogeneous linear ODE:
\begin{equation}
  M\frac{d^{2}y}{dt^{2}}+C\frac{dy}{dt}+Ky=0.
  \label{eq:eom}
\end{equation}
Assuming a solution of the form $e^{mt}$, the characteristic equation is
\begin{equation}
  Mm^{2}+Cm+K=0.
  \label{eq:char}
\end{equation}
with roots
\begin{equation}
  m_{1,2}=-\frac{C}{2M}\pm\frac{1}{2M}\sqrt{C^{2}-4MK}.
  \label{eq:roots}
\end{equation}

\subsection{Damping Regimes}
\label{sec:regimes}

\subsubsection{Overdamping ($C^{2}-4KM>0$)}
\label{sec:over}

The solution can be written as
\begin{equation}
  y(t)=e^{-(\frac{C}{2M})t}\!\left[c_{1}e^{\Omega t}+c_{2}e^{-\Omega t}\right],\quad
  \Omega=\frac{\sqrt{C^{2}-4KM}}{2M}.
  \label{eq:overdamped}
\end{equation}

\subsubsection{Critical Damping ($C^{2}=4KM$)}
\label{sec:critical}

There is a double root, giving
\begin{equation}
  y(t)=e^{-(\frac{C}{2M})t}\left[c_{1}+c_{2}t\right].
  \label{eq:critical}
\end{equation}

\subsubsection{Underdamping ($C^{2}-4KM<0$)}
\label{sec:under}

The most physically rich regime.  The solution is
\begin{equation}
  y(t)=e^{-\frac{C}{2M}t}\!\left[Ae^{i\Omega t}+Be^{-i\Omega t}\right],\quad
  \Omega=\frac{\sqrt{4KM-C^{2}}}{2M}.
  \label{eq:underdamped}
\end{equation}

\section{Lambert $W$ and the Underdamped Oscillator}
\label{sec:extrema_deriv}

We seek the extrema of \eqref{eq:underdamped}.
Differentiating and setting $dy/dt=0$, and using $e^{-Ct/(2M)}\neq0$,
\begin{equation}
  -\frac{C}{2M}\!\left(Ae^{i\Omega t}+Be^{-i\Omega t}\right)
  +i\Omega Ae^{i\Omega t}-i\Omega Be^{-i\Omega t}=0.
  \label{eq:extrema_cond}
\end{equation}
Rearranging:
\begin{equation}
  Ae^{i\Omega t}\!\left(-\tfrac{C}{2M}+i\Omega\right)
  =Be^{-i\Omega t}\!\left(\tfrac{C}{2M}+i\Omega\right),
  \label{eq:rearrange}
\end{equation}
which gives
\begin{equation}
  -\frac{B}{A}e^{-2i\Omega t}
  =\frac{\tfrac{C}{2M}-i\Omega}{\tfrac{C}{2M}+i\Omega}.
  \label{eq:ratio}
\end{equation}
Defining $\epsilon=-B/A$ and $\beta=C/(2M)$,
\begin{equation}
  \epsilon\, e^{-2i\Omega t}=\frac{\beta-i\Omega}{\beta+i\Omega}.
  \label{eq:eps}
\end{equation}

With the dimensionless variables
\begin{equation}
  u=\beta t,\qquad v=\Omega t,
  \label{eq:uv}
\end{equation}
equation~\eqref{eq:eps} becomes
\begin{equation}
  \epsilon\, e^{-2iv}=\frac{(u-iv)^{2}}{u^{2}+v^{2}}.
  \label{eq:squared}
\end{equation}
Note also that
\begin{equation}
  u^{2}+v^{2}=R^{2}=(\beta t)^{2}+(\Omega t)^{2}=(\omega_{0}t)^{2}.
  \label{eq:circle}
\end{equation}

\label{sec:lambert}

For a real-valued solution we require $B=\bar{A}$.
Writing $A=\tfrac{D}{2}e^{i\varphi}$ and $B=\tfrac{D}{2}e^{-i\varphi}$
(real amplitude $D$, initial phase $\varphi$),
\begin{equation}
  \frac{B}{A}=e^{-2i\varphi},\qquad
  \epsilon=-\frac{B}{A}=-e^{-2i\varphi}.
  \label{eq:BA}
\end{equation}

Substituting \eqref{eq:BA} into \eqref{eq:squared}:
\begin{equation}
  (u-iv)^{2}=-e^{-2i\varphi}R^{2}e^{-2iv}.
  \label{eq:sq2}
\end{equation}
Taking the square root (with sign ambiguity):
\begin{equation}
  u-iv=\pm\,i\,e^{-i\varphi}Re^{-iv}.
  \label{eq:sqrt}
\end{equation}
Multiplying both sides by $e^{i(v+\varphi)}$:
\begin{equation}
  (u-iv)\,e^{i(v+\varphi)}=\pm\,iR.
  \label{eq:rotated}
\end{equation}

Expanding $e^{i(v+\varphi)}$ in \eqref{eq:rotated} and separating real and imaginary parts:
\begin{align}
  \operatorname{Re}:&\quad u\cos(v+\varphi)+v\sin(v+\varphi)=0,
  \label{eq:re_cond}\\
  \operatorname{Im}:&\quad u\sin(v+\varphi)-v\cos(v+\varphi)=\pm R.
  \label{eq:im_cond}
\end{align}
Dividing \eqref{eq:re_cond} by $\cos(v+\varphi)$ (where non-zero) yields
the \emph{transcendental equation}(refer to figure 2a)

\begin{equation}
  {v\tan(v+\varphi)=-u.}
  \label{eq:transcendental}
\end{equation}

\begin{figure*}[h]
  \centering
  \includegraphics[width=\linewidth]{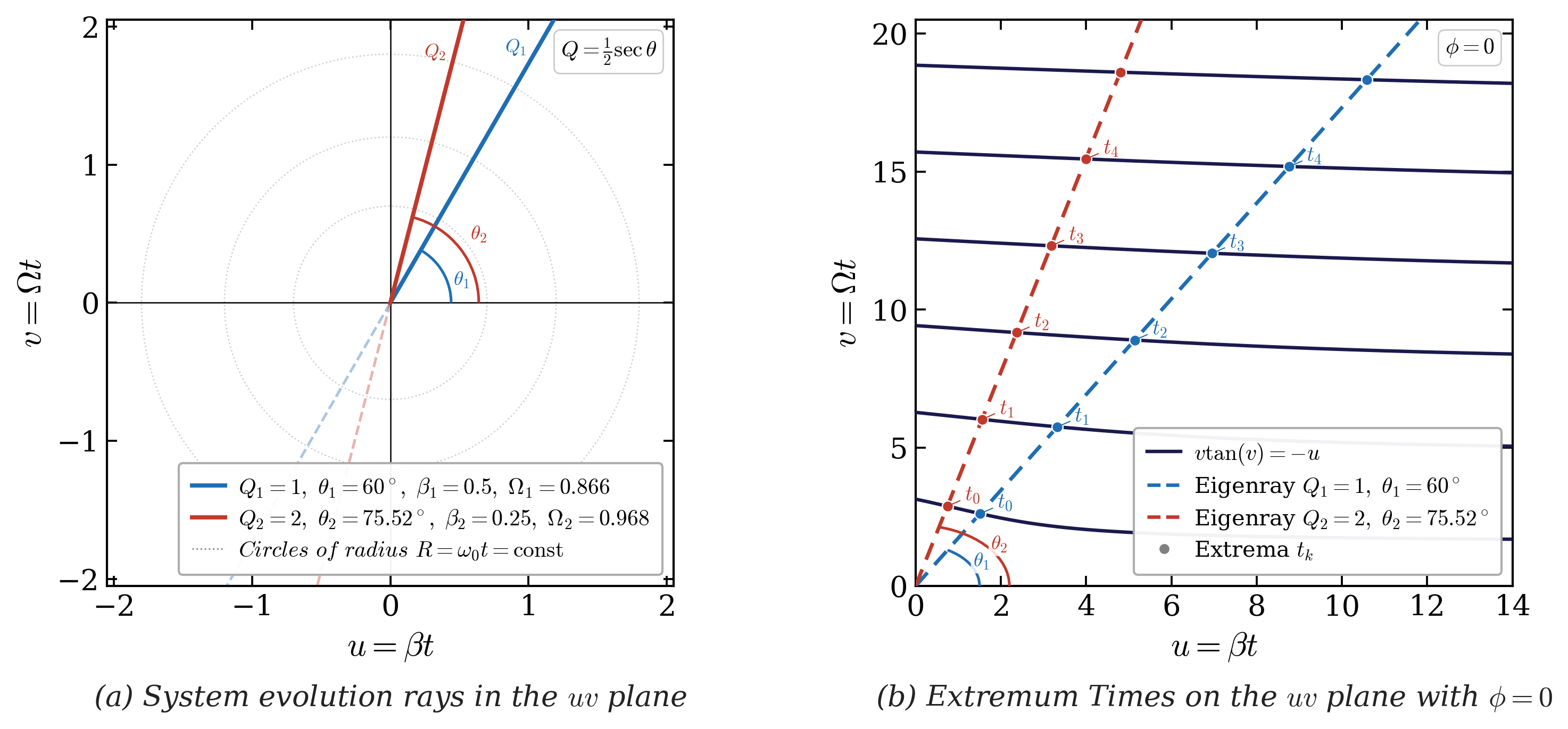}
  \caption{The black curves show the locus satisfying
    \eqref{eq:transcendental}.
    For $\varphi=0$ (left panel) these curves are \emph{not} the vertical
    axis: the equation $v\tan(v)=-u$ is satisfied only at the isolated
    points $(0,n\pi)$, not for all $v$ with $u=0$.
    The near-vertical appearance near $u=0$ is caused by $\tan(v)$ diverging
    at $v=(n+\tfrac{1}{2})\pi$.
    The red dashed ray traces the system's evolution; its intersections with
    the curves  give the extremum times $t_{0},t_{1},\ldots$}
  \label{fig:uv}
\end{figure*}

\subsubsection{The Rotated $z$-Plane Mapping}
\label{sec:zplane}

Condition~\eqref{eq:transcendental} emerges naturally from a single  rotation of the standard Lambert $W$ mapping in the $z$-plane\cite{roberts2017tutorial}.
Working directly with
\begin{equation}
  z=we^{-w},\qquad w=u+iv=\beta t+i\Omega t,
  \label{eq:zdef}
\end{equation}
and rotating by $-\varphi$:
\begin{equation}
  e^{-i\varphi}z=e^{-u}(u+iv)\,e^{-i(v+\varphi)}.
  \label{eq:zrot}
\end{equation}
Taking the real part of \eqref{eq:zrot} and using
\begin{equation}
    \operatorname{Re}[(u+iv)e^{-i(v+\varphi)}]=u\cos(v+\varphi)+v\sin(v+\varphi),
    \label{eq:re}
\end{equation}   
we get
\begin{equation}
  \operatorname{Re}(e^{-i\varphi}z)
  =e^{-u}\bigl[u\cos(v+\varphi)+v\sin(v+\varphi)\bigr].
  \label{eq:reZ}
\end{equation}
Setting $\operatorname{Re}(e^{-i\varphi}z)=0$ (with $e^{-u}\neq0$)
recovers precisely \eqref{eq:transcendental}.
Thus the extremum condition is equivalent to the rotated image $ze^{-i\varphi}$
lying on the imaginary axis in the $z$-plane.
Each time the ray at angle $\pi/2-\varphi$ in the $z$-plane is intersected
by a branch of $W$, we obtain a solution for $t$(refer to the figure 2b) .

\subsection{Geometry of the \textbf{$uv$} Plane}
\label{sec:uv}

The dimensionless variables $u=\beta t$ and $v=\Omega t$ define a plane
in which the system evolves along the ray
\begin{equation}
  u=\left(\frac{\beta}{\Omega}\right)v,
  \label{eq:ray}
\end{equation}
with slope $\tan\theta=v/u=\Omega/\beta$.\\\\
From \eqref{eq:circle}, $R=\sqrt{u^{2}+v^{2}}=\omega_{0}t$, so as time
increases the system moves outward along the ray, and a circle of
radius $R$ in the $uv$-plane expands.\\\\
From \eqref{eq:ray}, the slope of the evolution ray is
$\tan\theta=\Omega/\beta$.
For an underdamped oscillator, the quality factor:
\begin{equation}
  Q=\frac{\omega_{0}}{2\beta}
   =\frac{1}{2}\sqrt{1+\tan^{2}\theta}
   =\frac{1}{2}\sec\theta
  \label{eq:Q_theta} .
\end{equation}
\medskip
\subsection{Some explicit results for Underdamped Motion}
\subsubsection{Bounds on  $\theta$ and $Q$ }
For underdamped motion $\beta>0$ and $\Omega>0$, so $\tan\theta>0$ and
\begin{equation}
  0<\theta<\frac{\pi}{2},\qquad \frac{1}{2}<Q<\infty.
  \label{eq:range}
\end{equation}
These two inequalities are equivalent: one geometric, one dynamical.

\subsubsection{Limiting Cases}

\begin{itemize}
  \item \textbf{Light damping} ($\beta\ll\omega_{0}$):
        $\Omega\approx\omega_{0}$, so $\tan\theta\to\infty$,
        $\theta\to\pi/2$, and $Q\to\infty$.
        The ray approaches the $v$-axis; the oscillator rings many
        cycles before its amplitude decays.
  \item \textbf{Heavy damping approaching critical}
        ($\beta\to\omega_{0}^{-}$):
        $\Omega\to0^{+}$, $\theta\to0^{+}$, $Q\to1/2$.
        The ray collapses toward the $u$-axis; the oscillator barely
        completes one cycle.
\end{itemize}

\subsubsection{Threshold between Oscillatory and Aperiodic behaviour.}

\begin{figure}[H]
  \centering
  \includegraphics[width=\linewidth]{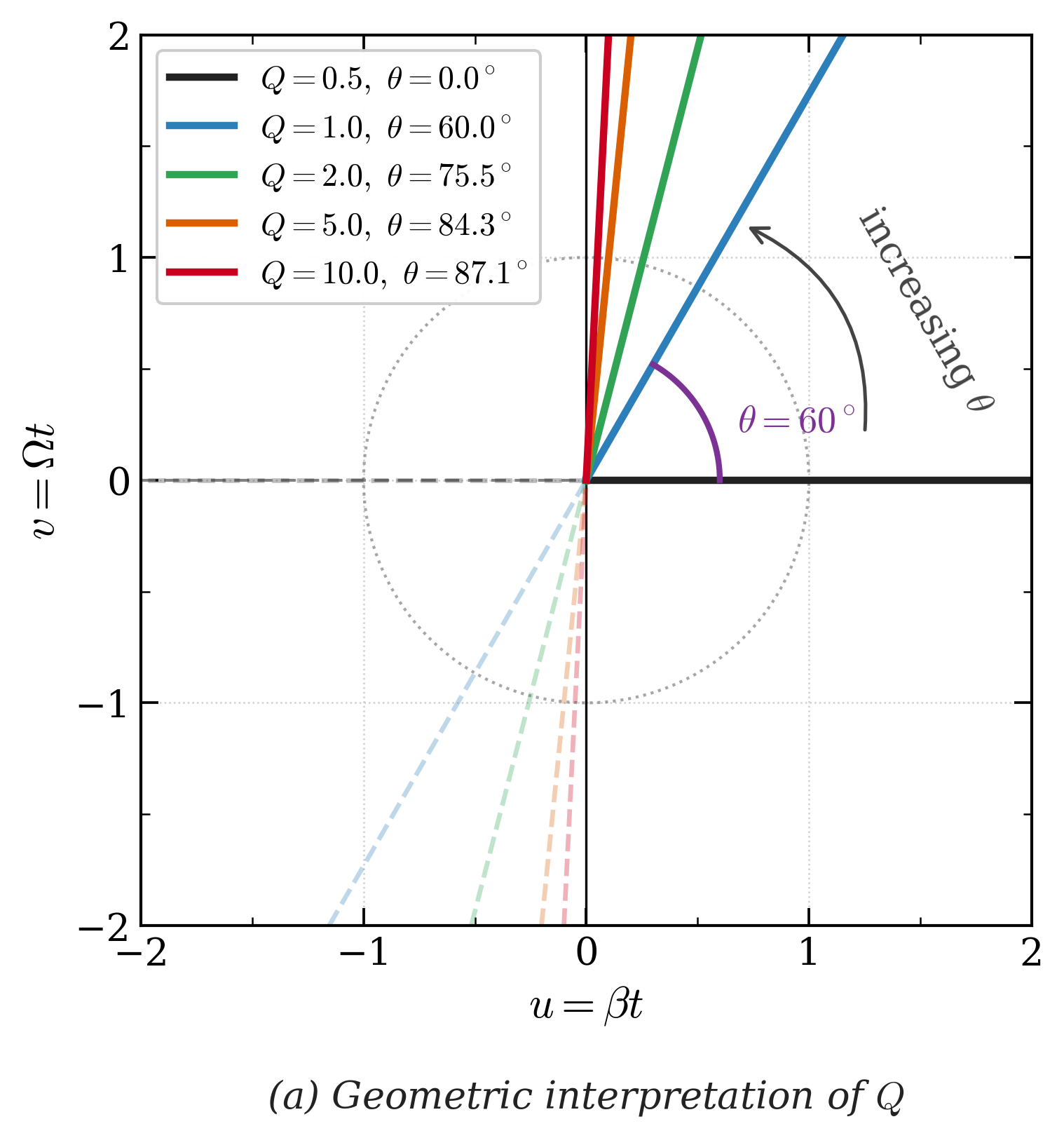}
  \caption{Geometric interpretation of $Q$ in the $uv$-plane
    ($u=\beta t$, $v=\Omega t$).
    Each ray corresponds to a fixed $Q$, with slope
    $\tan\theta=\Omega/\beta$ and $Q=\frac{1}{2}\sec\theta$.
    The horizontal ray ($\theta=0^{\circ}$) is the critically damped
    limit $Q=1/2$; as $\theta\to90^{\circ}$, $Q\to\infty$.
    Compare the ray angles here with the spiral figures:
    a larger $\theta$ produces more spiral turns.}
  \label{fig:Qrays}
\end{figure}

At $Q=1/2$: $\Omega=0$, $\theta=0$, ray on the $u$-axis.
This is the boundary between oscillatory and non-oscillatory behaviour.
For all underdamped and critically damped cases the $uv$-ray orientation
directly encodes $Q$ (Fig.~\ref{fig:Qrays}); we encourage the reader
to refer forward to Fig.~\ref{fig:Qrays} at this point.

\begin{figure*}[!ht]
  \centering
  \includegraphics[width=\linewidth]{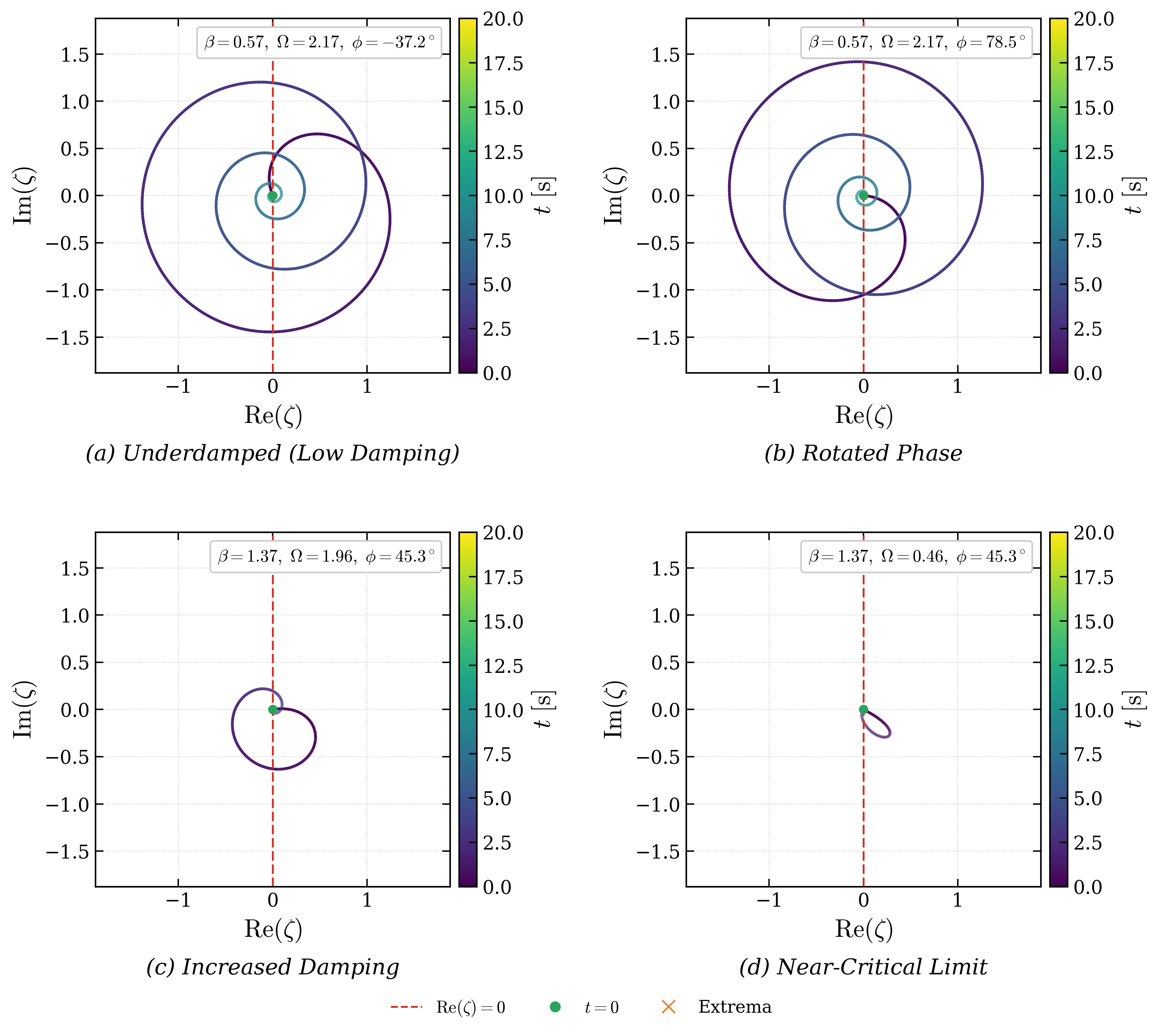}
  \caption{Spiral $\zeta(t)=e^{-i\varphi}we^{-w}$ for four underdamped
    parameter sets.
    Colour encodes time (colour bar at right); the red dashed line marks
    $\operatorname{Re}(\zeta)=0$ and its crossings give displacement extrema.
    \textbf{(a)}~$\beta=0.57$, $\Omega=2.17$, $\varphi=-37.2^\circ$:
    $Q\approx2.1$; spiral completes $\sim\!2$--$3$ turns with multiple
    imaginary-axis crossings visible.
    \textbf{(b)}~$\beta=0.57$, $\Omega=2.17$, $\varphi=78.5^\circ$:
    same $Q\approx2.1$ as~(a); changing $\varphi$ rotates the spiral
    without altering its shape or winding count.
    \textbf{(c)}~$\beta=1.37$, $\Omega=1.96$, $\varphi=45.3^\circ$:
    $Q\approx1.0$; faster decay produces fewer turns and only one or
    two axis crossings.
    \textbf{(d)}~$\beta=1.37$, $\Omega=0.46$, $\varphi=45.3^\circ$:
    $Q\approx0.6$, near the critically damped limit; barely half a turn
    completed.
    As $Q$ increases the spiral winds more times before contracting;
    changing $\varphi$ only rotates the pattern.}
  \label{fig:Underdamped}
\end{figure*}

\section{The \texorpdfstring{$\zeta$}{zeta}-Plane Spiral}
\label{sec:spiral}

\subsection{Orientation to the \texorpdfstring{$\zeta$}{zeta}-Plane}

The $\zeta$-plane is the complex plane~\cite{ablowitz2003} of the variable
\begin{equation}
  \zeta=e^{-i\varphi}we^{-w},\qquad w=\beta t+i\Omega t.
  \label{eq:zeta}
\end{equation}

Substituting $we^{-w}=(u+iv)e^{-u}e^{-iv}$ into \eqref{eq:zeta} and
expanding $e^{-i(v+\varphi)}$:
\begin{align}
  X(t)&=\operatorname{Re}(\zeta)
       =e^{-u}\bigl[u\cos(v+\varphi)+v\sin(v+\varphi)\bigr],
  \label{eq:X}\\
  Y(t)&=\operatorname{Im}(\zeta)
       =e^{-u}\bigl[v\cos(v+\varphi)-u\sin(v+\varphi)\bigr].
  \label{eq:Y}
\end{align}
These are the parametric equations of the spiral, with $u=\beta t$ and
$v=\Omega t$. The zeta plane is X,Y Plane

As time increases, $w$ traces a ray from the origin; the mapping
$t\mapsto\zeta(t)$ compresses this ray into a spiral that winds inward
as the oscillator decays.
Each point on the spiral corresponds to one instant in time (shown by
the colour bar in the figures).
Three key visuals are:
\begin{enumerate}
  \item \textbf{Number of loops} is directly proportional to the quality
        factor $Q$~\cite{kreyszig2011,thomson1998}. A high-$Q$ oscillator winds many times; a critically
        or over-damped system does not wind at all.
  \item \textbf{Imaginary-axis crossings} ($\operatorname{Re}(\zeta)=0$,
        red dashed line) -- each crossing marks a displacement extremum.
  \item \textbf{Spiral tightness} -- governed by $\beta$ (radial decay)
        and $\Omega$ (winding rate, the damped frequency).
\end{enumerate}

\subsection{Properties of the Spiral}
\label{sec:spiral_props}

\subsubsection{Radial coordinate.}
\begin{equation}
  |\zeta(t)|=|w|\,e^{-u}=\omega_{0}t\,e^{-\beta t}.
  \label{eq:radius}
\end{equation}
The radius is independent of $\varphi$; it grows from $0$ to a maximum
at $t_{\max}=1/\beta$ (found by $d|\zeta|/dt=0$), then decays
exponentially.

\subsubsection{Extremum Condition}

Setting $X(t)=\operatorname{Re}(\zeta)=0$ in \eqref{eq:X} yields
exactly \eqref{eq:transcendental}: the extremum times are the instants
when $\zeta(t)$ crosses the imaginary axis.

Starting from \eqref{eq:transcendental} with $u=\beta t$, $v=\Omega t$,
and substituting $u=(\beta/\Omega)v$:
\begin{equation}
  \tan(v+\varphi)=-\frac{\beta}{\Omega}=-\cot\theta=\tan\!\left(\theta-\frac{\pi}{2}\right).
  \label{eq:tan_cond}
\end{equation}
The general solution of \eqref{eq:tan_cond} is
$v+\varphi=\theta-\pi/2+n\pi$, $n\in\mathbb{Z}$, giving
\begin{equation}
  {t_{n}=\frac{\theta-\varphi-\frac{\pi}{2}+n\pi}{\Omega},
  \quad n=0,1,2,\ldots}
  \label{eq:tn}
\end{equation}
(retaining only $t_{n}\geq0$).  Successive extrema are separated by
$\Delta t=\pi/\Omega$, consistent with the oscillation period
$T=2\pi/\Omega$.

\subsection{Q Factor }   
    The quality factor \( Q = \omega_0 / (2\beta) = \frac{1}{2} \sec \theta \) determines how many turns the spiral makes before its radius decays significantly. For large \( Q \) (small \( \beta \)), \( \theta \approx \pi / 2 \) and the spiral winds many times; for \( Q \) just above \( 1/2 \), it winds only a fraction of a turn.

    \subsection{Winding rate}
    The spiral \(\zeta(t) = e^{-i\phi}we^{-w}\) with \(w = \beta t + i\Omega t\) has an argument

\[\arg \zeta(t) = \arctan \left( \frac{\Omega}{\beta} \right) - \Omega t - \phi = \theta - \Omega t - \phi,\]

where \(\theta = \arctan(\Omega / \beta)\) is the fixed ray angle. Thus the \textbf{angular speed} (winding rate) is

\[\frac{d}{dt} \arg \zeta(t) = -\Omega,\]

a negative constant.
Therefore, \(\Omega\) \textbf{directly determines how fast the spiral winds}: larger \(\Omega\) gives faster rotation (more turns per unit time). The magnitude of \(\Omega\) is the damped frequency of the oscillator, so the spiral completes one full revolution in time \(2\pi / \Omega\), exactly the period of the damped oscillation
   The spiral \( \zeta(t) = e^{-i\phi} w e^{-w} \) is a compact geometric representation of the underdamped harmonic oscillator. It encapsulates all physical parameters: \( \beta \) governs the radial decay, \( -\Omega \) the winding rate, \( \phi \) the orientation, and \( Q \) the number of turns. \\

   \begin{figure}
    \centering
    \includegraphics[width=1.00\linewidth]{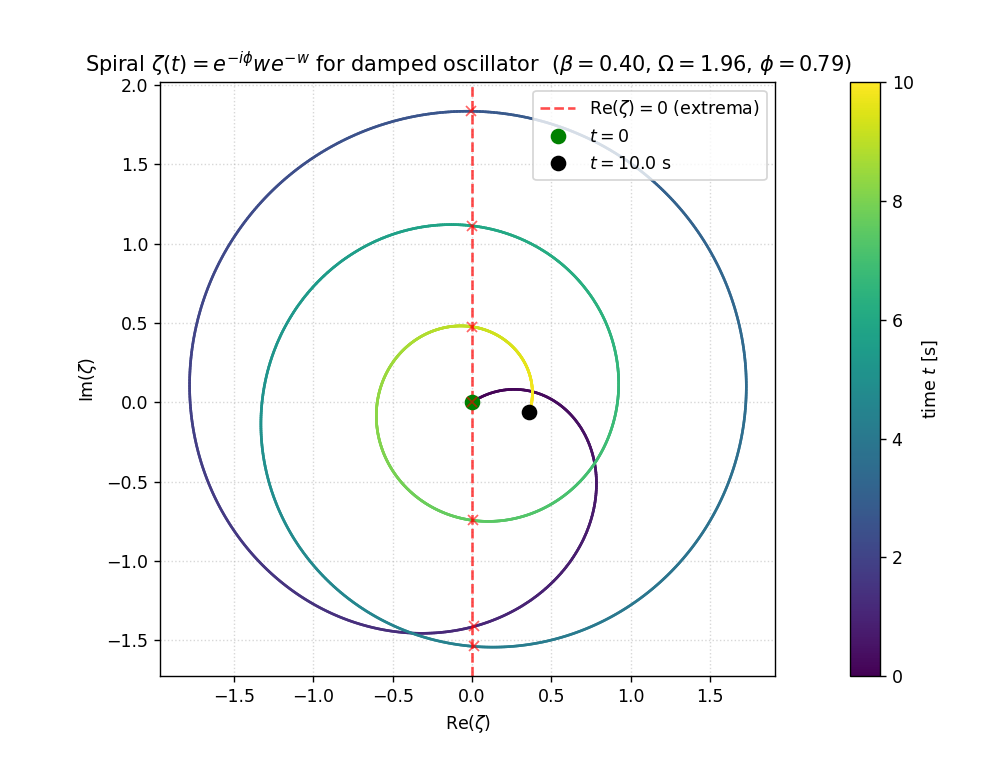}
    \caption{A damped oscillator with time stamp from 0 to 10 secs . The green dot indicates  the start of the motion of spiral and black dot indicates the end. }
    \label{fig:placeholder}
\end{figure}

\subsubsection{\textbf{Lambert W Solution for Spiral Radius Threshold}}
\label{subsec:lambert_threshold}

The magnitude of the spiral variable  $\zeta(t)=e^{-i\varphi}we^{-w}$ is (from equation \eqref{eq:radius}),
\[
\zeta(t)| = \omega_0 t\, e^{-\beta t},
\]

where $\omega_0=\sqrt{K/M}$ is the undamped natural frequency and
$\beta=C/(2M)$ is the damping coefficient.
We ask: \emph{At what times does the spiral cross a circle of radius
$A$ in the complex plane?}
That is solve
\begin{equation}
\omega_0 t\, e^{-\beta t}=A,
\label{eq:radius_threshold}
\end{equation}
where $0<A<\max|\zeta|=\omega_0/(\beta e)$ (otherwise no real
positive solution exists).

Dividing equation \eqref{eq:radius_threshold} by $\omega_0$ and setting
$x=\beta t$ (so $t=x/\beta$):
\[
\frac{x}{\beta}e^{-x}=\frac{A}{\omega_0}
\quad\Longrightarrow\quad
x e^{-x}= \frac{\beta A}{\omega_0}\equiv c.
\]
Multiplying by $-1$ and writing $u=-x$ gives
\[
u e^{u}=-c,
\]
which is the defining equation of the Lambert $W$ function:
$u=W_k(-c)$. Hence
\[
-x = W_k\!\left(-\frac{\beta A}{\omega_0}\right)
\quad\Longrightarrow\quad
x = -W_k\!\left(-\frac{\beta A}{\omega_0}\right).
\]
Returning to $t=x/\beta$ yields the closed‑form expression
\begin{equation}
{t = -\frac{1}{\beta}\,W_k\!\left(-\frac{\beta A}{\omega_0}\right),\qquad
k\in\{0,-1\}.}
\label{eq:lambert_t}
\end{equation}

The argument $-\beta A/\omega_0$ lies in $[-1/e,0)$ because
$A\le\omega_0/(\beta e)$. For this interval the Lambert $W$
function has two real branches:
\begin{itemize}
\item The principal branch $W_0$ gives values in $[-1,0)$,
  leading to $x\in(0,1]$ and therefore a \emph{small} time
  $t_{\text{rise}}$ (the spiral is still expanding toward its
  maximum radius).
\item The lower branch $W_{-1}$ gives values in $(-\infty,-1]$,
  yielding $x\in[1,\infty)$ and a \emph{larger} time
  $t_{\text{decay}}$ (the spiral has passed its maximum and is
  shrinking).
\end{itemize}
When $A=|\zeta|_{\max}=\omega_0/(\beta e)$, both branches meet
at $x=1$, giving a single solution $t=1/\beta$.

\subsubsection{Winding Number.}
The quality factor for an underdamped oscillator,
\begin{equation}
  Q=\frac{1}{2}\sec\theta
  \label{eq:Q_theta_dup}
\end{equation}
determines how many turns the spiral makes before its radius decays significantly. For large $Q$ (small $\beta$), $\theta\approx\pi/2$ and the spiral
winds many times; for $Q$ just above $1/2$, it barely completes a
single turn.\\
Thus the angle $\theta$ of the evolution ray in the $uv$-plane
\emph{uniquely} determines $Q$: fixing $\theta$ fixes all global
dynamics (see Fig.~\ref{fig:Qrays}).\\

The argument of $\zeta(t)$ is
\begin{equation}
  \arg\zeta(t)=\arctan\!\left(\frac{\Omega}{\beta}\right)-\Omega t-\varphi
              =\theta-\Omega t-\varphi,
  \label{eq:arg}
\end{equation}
so the angular speed is $d(\arg\zeta)/dt=-\Omega$, a negative constant.
Thus $\Omega$ directly determines the winding rate: the spiral completes
one full revolution in time $2\pi/\Omega$, exactly the period of the
damped oscillation.
\\The total change in argument from \( t = 0 \) to \( t = T \) is
\[
\Delta \arg \zeta = -\Omega T.
\]

The winding number\cite{brown2009} (number of counter-clockwise turns around the origin) is
\[
\text{Winding}(T) = \frac{\Delta \arg \zeta}{2\pi} = -\frac{\Omega T}{2\pi}.
\]

The negative sign indicates clockwise winding. As \( T \to \infty \), the winding number diverges to \(-\infty\); mathematically, the spiral winds infinitely many times as its radius $|\zeta(t)|=\omega_0 t e^{-\beta t}$ decays exponentially. Let us  define an effective winding number $N_\varepsilon$ as the number of turns completed before the spiral radius falls below a small threshold $\varepsilon>0$:
\[
|\zeta(t_\varepsilon)| = \varepsilon \quad\Longrightarrow\quad \omega_0 t_\varepsilon e^{-\beta t_\varepsilon} = \varepsilon.
\] 

We require $t_\varepsilon>0$ such that
\[
\omega_0 t_\varepsilon e^{-\beta t_\varepsilon} = \varepsilon.
\]
Set $x=\beta t_\varepsilon$.  Then
\[
\frac{\omega_0}{\beta}\,x e^{-x} = \varepsilon \quad\Longrightarrow\quad x e^{-x} = \frac{\beta\varepsilon}{\omega_0}.
\]
Multiplying by $-1$ and letting $u=-x$ gives $u e^{u} = -\beta\varepsilon/\omega_0$.  Hence
\[
u = W_k\!\left(-\frac{\beta\varepsilon}{\omega_0}\right),\qquad
x = -W_k\!\left(-\frac{\beta\varepsilon}{\omega_0}\right).
\]
For $0<\varepsilon<\max|\zeta|$, the argument $-\beta\varepsilon/\omega_0$ lies in $(-1/e,0)$.  In this interval the equation $x e^{-x}=c$ has two positive solutions, corresponding to the two real branches of the Lambert $W$ function:
\[
x_{\text{small}} = -W_0(-c), \qquad x_{\text{large}} = -W_{-1}(-c),
\]
where $c=\beta\varepsilon/\omega_0$.  The small solution (using $W_0$) gives the time when the spiral first reaches radius $\varepsilon$ on its outward (expanding) part; the large solution (using $W_{-1}$) gives the later time when the spiral, after having passed its maximum radius, decays back to $\varepsilon$.  For the effective winding number we count the total number of turns up to the moment the spiral finally shrinks below $\varepsilon$, i.e., we take the larger time:
\[
t_\varepsilon = \frac{x_{\text{large}}}{\beta} = -\frac{1}{\beta}\,W_{-1}\!\left(-\frac{\beta\varepsilon}{\omega_0}\right).
\]

Over one decay time \( T_{\text{decay}} = 1/\beta \), the number of clockwise turns is
\begin{equation}
N_{\text{turns}} = \bigl|\text{Winding}(1/\beta)\bigr| = \frac{\Omega}{2\pi\beta}
\label{eq:N_turns}
\end{equation}

Now, the quality factor \( Q \) for an underdamped oscillator (with decay rate \( \beta \) and damped frequency \( \Omega \)) is defined as
\[
Q = \frac{\omega_0}{2\beta}, \quad \text{where} \quad \omega_0 = \sqrt{\beta^2 + \Omega^2}.
\]
Thus
\begin{equation}
Q = \frac{\sqrt{\beta^2 + \Omega^2}}{2\beta}
= \frac{1}{2}\sqrt{1 + \left(\frac{\Omega}{\beta}\right)^2}
\label{eq:Q}
\end{equation}

From \eqref{eq:N_turns}, \( \frac{\Omega}{\beta} = 2\pi N_{\text{turns}} \). Substituting into \eqref{eq:Q} gives

\begin{equation}
Q = \frac{1}{2}\sqrt{1 + \left(2\pi N_{\text{turns}}\right)^2}
\label{eq:Q_N_turns}
\end{equation}

The spiral $\zeta(t)=e^{-i\varphi}we^{-w}$ is a compact geometric
representation of the underdamped harmonic oscillator: $\beta$ governs
the radial decay, $\Omega$ the winding rate, $\varphi$ the orientation,
and $Q$ the number of turns\\
\subsubsection*{Asymptotic Winding Number for Large $Q$}
 
For large $Q$, the effective winding number $N_\varepsilon$
admits a closed-form asymptotic expression via the
$W_{-1}$ branch.
\[
N_\varepsilon = \frac{\Omega t_\varepsilon}{2\pi}
= -\frac{\Omega}{2\pi\beta}\,W_{-1}\!\left(-\frac{\beta\varepsilon}{\omega_0}\right).
\]
Recall $\beta = \omega_0/(2Q)$, so the argument of $W_{-1}$
satisfies
\[
  y \;=\; \frac{\beta\varepsilon}{\omega_0} \;=\; \frac{\varepsilon}{2Q} \;\to\; 0^+
  \quad \text{as } Q\to\infty.
\]
Using the standard asymptotic expansion\cite{corless1996}
$W_{-1}(-y) = \ln y - \ln(-\ln y) + o(1)$ as $y\to0^+$,
and noting $-\ln y = \ln(2Q/\varepsilon)$, gives
\[
  x_\text{large} \;=\; -W_{-1}(-y)
  \;=\; \ln\!\left(\tfrac{2Q}{\varepsilon}\right)
        + \ln\!\ln\!\left(\tfrac{2Q}{\varepsilon}\right)
        + o(1).
\]
Combining with $\Omega/\beta = \sqrt{4Q^2-1}
= 2Q - 1/(4Q) + O(Q^{-3})$ and
$N_\varepsilon = (\Omega/2\pi\beta)\,x_\text{large}$,
the leading-order result is
\begin{equation}
  {N_\varepsilon \;\approx\;
  \frac{Q}{\pi}\,\ln\!\left(\frac{2Q}{\varepsilon}\right)},
  \qquad Q \gg 1,
  \label{eq:Neps_asymp}
\end{equation}
with the two-term refinement
\begin{equation}
  N_\varepsilon \;=\;
  \frac{Q}{\pi}
  \left[
    \ln\!\left(\tfrac{2Q}{\varepsilon}\right)
    + \ln\!\ln\!\left(\tfrac{2Q}{\varepsilon}\right)
  \right]
  + O(1).
  \label{eq:Neps_twoterm}
\end{equation}
a high‑$Q$ oscillator completes many turns before its spiral becomes indistinguishable.  Moreover, the formula provides a practical way to estimate $Q$ from the number of visible oscillations in an experimental signal.

The Mössbauer effect\cite{bloch2013introduction} involves the resonant absorption of gamma rays by atomic nuclei bound in a crystal lattice. This nuclear resonance has an extraordinarily high quality factor \(Q\) – far larger than typical mechanical or electrical oscillators. In fact, the Mössbauer resonance is one of the highest‑\(Q\) systems known in physics. The geometric framework developed in this paper for the damped harmonic oscillator centers on the quality factor as a measure of how many oscillations occur before energy decays. That same idea applies directly to the Mössbauer effect: a high \(Q\) means the nuclear excitation persists for many cycles, producing a sharp energy line. 
\subsection{Determining $Q$ from the spiral: methods and
comparison}
 
Three geometric methods for measuring $Q$ from the
spiral $\zeta(t)$ are now available.
Each exploits a different aspect of the spiral's structure
and is suited to a different experimental context.
The ray angle in the complex $uv$ plane is
geometrically exact, and gives $Q$ via the
formula
\begin{equation}
	Q=\frac{1}{2}\sec\theta .
	\label{eq:Q_theta_dup2}
\end{equation}
It is theoretically correct, but sometimes
experimentally inconvenient.

Besides the ray angle in the complex plane, 
we can use the methods of logarithmic decrement
or spiral turn counting.
Here we detail those two methods.
 
\subsubsection{Logarithmic decrement}

The classical method records successive displacement peaks $y_n$ and
forms the logarithmic decrement~\cite{bloch2013introduction}
\[
\delta = \ln\!\left(\frac{y_n}{y_{n+1}}\right)
= \frac{2\pi\beta}{\Omega}.
\]
The quality factor follows from
\begin{equation}
Q = \frac{1}{2}\sqrt{1 + \left(\frac{2\pi}{\delta}\right)^{2}}
\;\approx\; \frac{\pi}{\delta}, \qquad Q \gg 1.
\end{equation}
This requires resolving individual amplitude peaks, which
becomes unreliable for very high-$Q$ systems where
successive peaks differ by a tiny fraction.

\subsubsection{Spiral turn counting}
 
An alternative is to count the number of complete oscillations $N_\varepsilon$
visible above a threshold $\varepsilon$ (as a fraction of
$\max|\zeta|$), then solve~\eqref{eq:Neps_asymp}
for $Q$:
\begin{equation}
  {N_\varepsilon \;\approx\;
  \frac{Q}{\pi}\,\ln\!\left(\frac{2Q}{\varepsilon}\right)},
  \qquad Q \gg 1,
  \label{eq:Q_turns}
\end{equation}
For high-$Q$ systems this method is advantageous:
counting oscillations is robust to noise, whereas resolving
the amplitude ratio of closely spaced peaks (as in the log
decrement) is noise-sensitive.

The presence of the logarithm inside the product \(Q \ln(2Q/\epsilon)\) makes direct inversion impossible with elementary functions. However, the Lambert \(W\) function—defined as the multi‑branch inverse of \(w e^{w} = z\)—provides an elegant closed‑form solution.
Start from the asymptotic relation for a high‑\(Q\) oscillator:

\[
N_{\epsilon} = \frac{Q}{\pi}\,\ln\!\left(\frac{2Q}{\epsilon}\right), \qquad Q \gg 1. \tag{45}
\]

Introduce  the dimensionless variables

\[
x = \frac{Q}{\pi}, \qquad K = \frac{2\pi}{\epsilon},
\]

so that \(2Q/\epsilon = Kx\) and (45) becomes

\[
N_{\epsilon} = x \ln(Kx).
\]

Multiply  both sides by \(K\):

\[
K N_{\epsilon} = Kx \ln(Kx).
\]

Set \(y = Kx\); then

\[
y \ln y = K N_{\epsilon}.
\]

Now write  \(y = e^{\ln y}\). Substituting gives \((\ln y)\, e^{\ln y} = K N_{\epsilon}\).  
By definition of the Lambert \(W\) function, \(W(z)e^{W(z)} = z\), we identify

\[
\ln y = W(K N_{\epsilon}) \quad\Longrightarrow\quad y = e^{W(K N_{\epsilon})}.
\]

Returning to the original variables via \(y = Kx = \frac{2\pi}{\epsilon}\cdot\frac{Q}{\pi} = \frac{2Q}{\epsilon}\), we obtain

\[
\frac{2Q}{\epsilon} = e^{\,W(K N_{\epsilon})}
\quad\Longrightarrow\quad
Q = \frac{\epsilon}{2}\; \exp\!\Bigl[W\!\bigl(K N_{\epsilon}\bigr)\Bigr].
\]

Finally, substituting \(K = 2\pi/\epsilon\) yields the closed‑form inversion

.
\begin{equation}
   Q = \frac{\epsilon}{2}\; \exp\!\left[ W\!\left( \frac{2\pi N_{\epsilon}}{\epsilon} \right) \right].
   \label{eq:Q_value}
\end{equation}

\subsubsection{Determining Q values Numerically}
\subsubsection*{Comparison of Methods}
\begin{enumerate}

  \item \textbf{Ray angle in the $uv$-plane} -- Exact; uses the slope of the 
  line $(u, v) = (\beta t, \Omega t)$ 
  to compute 
    \[
    Q = \frac{1}{2}\sec\theta.
    \]

    \item \textbf{Log decrement} -- Measures the exponential decay of successive displacement peaks, and estimates $Q$ using
    \begin{equation}
    	Q \approx \frac{1}{2}\sqrt{1 + \left(\frac{2\pi}{\delta}\right)^{2}}
    	\;\approx\; \frac{\pi}{\delta}, \qquad Q \gg 1.
    \end{equation}

    \item \textbf{Spiral turn counting} -- Counts the number of visible oscillations $N_\varepsilon$ before the spiral radius falls below a threshold, and estimates $Q$ using
    \begin{equation}
    	Q \approx \frac{\epsilon}{2}\; \exp\!\left[ W\!\left( \frac{2\pi N_{\epsilon}}{\epsilon} \right) \right], \qquad Q \gg 1.
    \end{equation}
    
\end{enumerate}
 

\begin{table}[htbp]
\centering
\caption{Numerical verification of the damped oscillator spiral mapping. The ray-angle method \(Q_{\text{ray}} = \frac{1}{2}\sec\theta\), the turn-counting method \(Q_{\text{turns}}\) (solving \(N_{\epsilon} = \frac{Q}{\pi}\ln(2Q/\epsilon)\)), and the logarithmic decrement method \(Q_{\text{logdec}}\) are compared. The effective number of turns \(N_{\epsilon}\) and the threshold \(\epsilon\) (as a fraction of \(\max|\zeta|\)) are also listed.}
\label{tab:numerical_verification}
\begin{tabular}{cccccc}
\toprule
\(Q_{\text{ray}}\) & \(Q_{\text{turns}}\) & \(Q_{\text{logdec}}\) & \(N_{\epsilon}\) & \(\epsilon\) \\
\midrule
2.50 & 2.31 & 2.48 & 2.10 & 0.0602 \\
5.00 & 4.92 & 4.98 & 5.28 & 0.0426 \\
10.00 & 9.94 & 9.97 & 12.59 & 0.0341 \\
25.00 & 24.86 & 24.92 & 37.22 & 0.0263 \\
\bottomrule
\end{tabular}
\end{table}

\subsection{Area Enclosed by the Spiral $\zeta(t)$}

For a parametric curve in the complex plane, Green's theorem gives:\cite{needham2023}(Chapter 11, section 2)
\begin{equation}
    A = \frac{1}{2}\left|\int_0^T \operatorname{Im}(\bar{\zeta}\,\dot{\zeta})\,dt\right|
\end{equation}

Starting from $\zeta = e^{-i\varphi}we^{-w}$, direct differentiation gives
\begin{equation}
    \dot{\zeta} = e^{-i\varphi}(1-w)\dot{w}\,e^{-w},
\end{equation}
and taking the complex conjugate 
of $\zeta$ gives
\begin{equation}
    \bar{\zeta} = e^{+i\varphi}\bar{w}\,e^{-\bar{w}}.
\end{equation}
Forming the product $\bar{\zeta}\dot{\zeta}$,
\begin{equation}
    \bar{\zeta}\,\dot{\zeta} = \bar{w}(1-w)\dot{w}\,e^{-(\bar{w}+w)}.
\end{equation}
Since $w = (\beta + i\Omega)t$, we have $\bar{w} + w = 2\operatorname{Re}(w) = 2\beta t$
and $\dot{w} = \beta + i\Omega$. Substituting:
\begin{equation}
    \bar{w}(1-w)\dot{w}
    = (\beta - i\Omega)t\,\bigl(1 - (\beta+i\Omega)t\bigr)(\beta+i\Omega).
\end{equation}
Expanding and using $(\beta - i\Omega)(\beta + i\Omega) = \beta^2 + \Omega^2$:
\begin{equation}
    \bar{w}(1-w)\dot{w}
    = (\beta^2+\Omega^2)\bigl[t - (\beta+i\Omega)t^2\bigr].
\end{equation}
Taking the imaginary part, and noting that $t$ is real so only the
$i\Omega t^2$ term contributes:
\begin{equation}
    \operatorname{Im}(\bar{\zeta}\,\dot{\zeta})
    = -(\beta^2+\Omega^2)\,\Omega\,t^2\,e^{-2\beta t}.
\end{equation}

The area integral, 

\begin{equation}
    A = \frac{(\beta^2+\Omega^2)\,\Omega}{2}\int_0^T t^2\, e^{-2\beta t}\,dt.
\end{equation}
Integrating by parts twice:
\begin{equation}
    \int_0^T t^2 e^{-2\beta t}\,dt = \frac{1}{4\beta^3} 
    - e^{-2\beta T}\!\left(\frac{T^2}{2\beta}+\frac{T}{2\beta^2}+\frac{1}{4\beta^3}\right)
\end{equation}

The total Area of the full spiral is evaluated by taking the limit $T\to\infty$, and it gives

\begin{equation}
    \int_0^\infty t^2 e^{-2\beta t}\,dt = \frac{1}{4\beta^3}.
\end{equation}
Thus we have the closed-form result for the full area:
\begin{equation}
{A = \frac{(\beta^2+\Omega^2)\,\Omega}{8\beta^3}}
\end{equation}

Using $\beta^2 + \Omega^2 = \omega_0^2$ 
\begin{equation}
    A = \frac{\omega_0^2\,\Omega}{8\beta^3}
\end{equation}
    In the lightly damped limit $\beta \ll \Omega$:
\[
A \approx \frac{\Omega^3}{8\beta^3}
= \frac{1}{8}\left(\frac{\Omega}{\beta}\right)^3
= Q^3 \quad
\]
so higher-$Q$ systems sweep vastly more area before decaying.

\section{Spirals Across the Damping Regimes}
\subsection{Underdamped Regime ($Q>1/2$)}
\label{sec:underdamped}

The quality factor is $Q=\omega_{0}/(2\beta)$.  Expressing $\beta$ and
$\Omega$ in terms of $Q$:
\begin{equation}
  \beta=\frac{\omega_{0}}{2Q},\qquad
  \Omega=\omega_{0}\sqrt{1-\frac{1}{4Q^{2}}}.
  \label{eq:Q_params}
\end{equation}
Introduce the dimensionless time $\tau=\omega_{0}t$:
\begin{equation}
  u=\frac{\tau}{2Q},\qquad v=\tau\sqrt{1-\frac{1}{4Q^{2}}}.
  \label{eq:tau}
\end{equation}
The spiral radius becomes
\begin{equation}
  R(\tau)=|\zeta|=\tau\,e^{-\tau/(2Q)},
  \label{eq:R_tau}
\end{equation}
reaching its maximum at $\tau=2Q$ (i.e.\ $t=2Q/\omega_{0}$):
\begin{equation}
  R_{\max}=\frac{2Q}{e}.
  \label{eq:Rmax}
\end{equation}
A higher $Q$ gives a larger initial outward swing before contraction
dominates.

Figure~\ref{fig:Underdamped} shows four underdamped spirals for different
parameter sets.  The panels illustrate how $Q$ (winding count) and
$\varphi$ (rotation) independently control the spiral's appearance.


\subsection{Critically Damped Regime ($Q=1/2$)}
\label{sec:crit}

Here $\beta=\omega_{0}$ and $\Omega=0$.  The variable $w$ becomes
purely real:
\begin{equation}
  w=\omega_{0}t,\quad v=0
  \quad\Rightarrow\quad
  \zeta=e^{-i\varphi}\omega_{0}t\,e^{-\omega_{0}t}.
  \label{eq:crit_zeta}
\end{equation}
The spiral degenerates to a segment on the real axis (rotated by
$\varphi$).  No oscillation occurs; no imaginary-axis crossings exist.

\subsection{Overdamped Regime ($Q<1/2$)}
\label{sec:over2}

Now $\beta>\omega_{0}$ and $\Omega=i\gamma$ where
$\gamma=\sqrt{\beta^{2}-\omega_{0}^{2}}$.
The variable $w=(\beta-\gamma)t$ is real and positive, so $\zeta$ maps
to a real curve (rotated by $e^{-i\varphi}$).
No winding occurs; the motion is aperiodic.

\begin{figure*}[!ht]
  \centering
  \includegraphics[width=\linewidth]{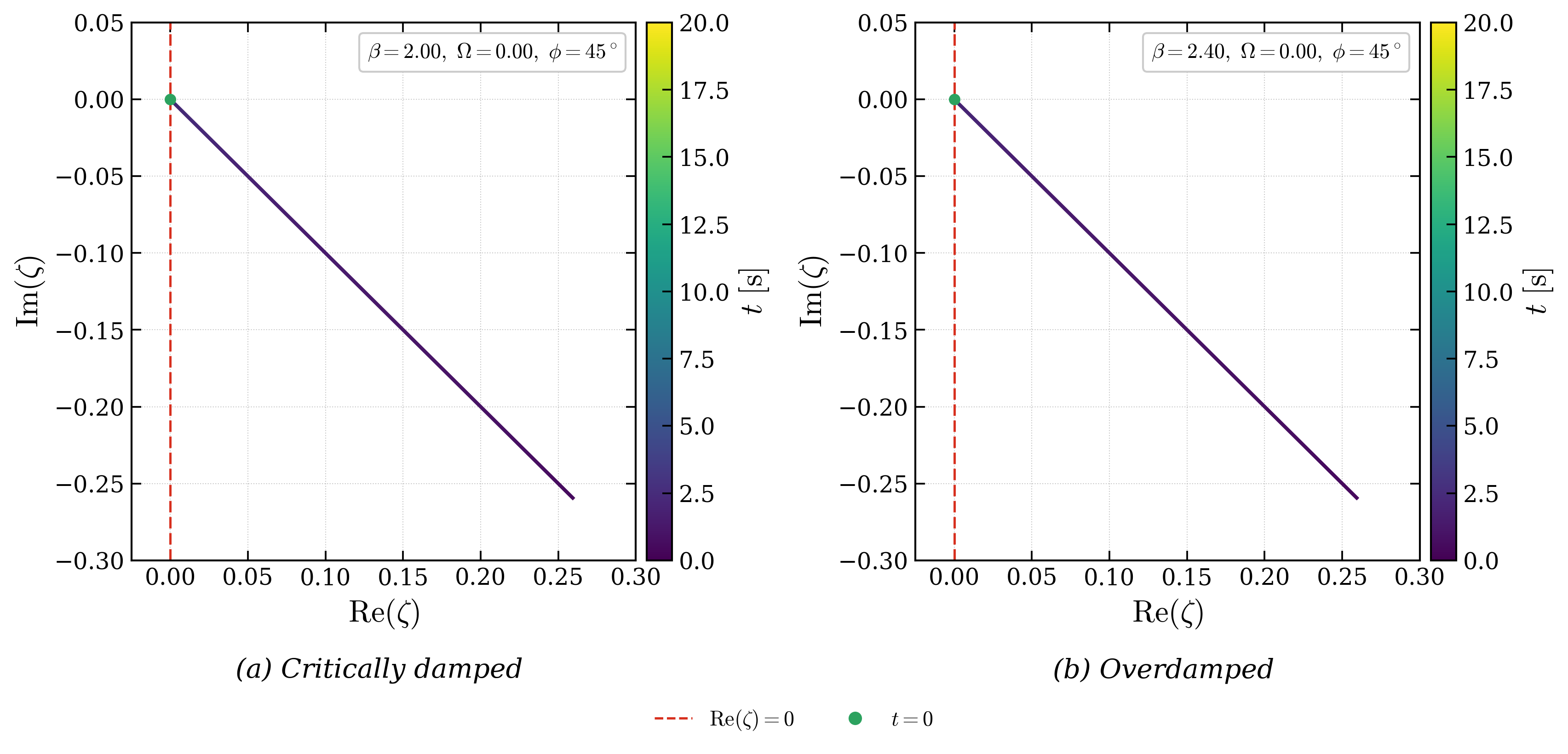}
  \caption{Spiral $\zeta(t)=e^{-i\varphi}we^{-w}$ for the
    critically damped ($Q=0.5$), and overdamped
    ($Q=0.3$) harmonic oscillator ($\varphi=\pi/4$).
    Colour encodes time.
    The red dashed line marks $\operatorname{Re}(\zeta)=0$.
    In the critically and over-damped cases the spiral collapses to a
    straight line, confirming aperiodic motion.}
  \label{fig:regimes}
\end{figure*}

\section{Energy of an Underdamped Oscillator}
\label{sec:energy}

\subsection{Displacement and Velocity}

The displacement is
\begin{equation}
  y(t)=De^{-\beta t}\cos(\Omega t+\varphi),
  \label{eq:disp}
\end{equation}
where $\beta=C/2M$, $\Omega=\sqrt{\omega_{0}^{2}-\beta^{2}}$,
$\omega_{0}=\sqrt{K/M}$.
Let $\theta_{\!s}\equiv\Omega t+\varphi$ for conciseness.
The velocity is
\begin{equation}
  \dot{y}=-De^{-\beta t}(\beta\cos\theta_{\!s}+\Omega\sin\theta_{\!s}).
  \label{eq:vel}
\end{equation}

\subsection{Total Mechanical Energy}

$E=\tfrac{1}{2}M\dot{y}^{2}+\tfrac{1}{2}Ky^{2}$.
Substituting \eqref{eq:disp}--\eqref{eq:vel}:
{\small
\begin{align}
  E(t)=\tfrac{1}{2}MD^{2}e^{-2\beta t}\Bigl[
  &\underbrace{\beta^{2}c_{\theta}^{2}+2\beta\Omega c_{\theta}s_{\theta}
        +\Omega^{2}s_{\theta}^{2}}_{\text{from }\dot{y}^{2}}
  +\underbrace{\tfrac{K}{M}c_{\theta}^{2}}_{\text{from }y^{2}}\Bigr],
  \label{eq:energy_full}
\end{align}
}%
where $c_{\theta}=\cos\theta_{\!s}$, $s_{\theta}=\sin\theta_{\!s}$.

Grouping the $\cos^{2}\theta_{\!s}$ coefficient:
$\beta^{2}+K/M=\beta^{2}+\omega_{0}^{2}$.
Applying double-angle identities and using $\Omega^{2}=\omega_{0}^{2}-\beta^{2}$
(so $\beta^{2}+\omega_{0}^{2}+\Omega^{2}=2\omega_{0}^{2}$ and
$\beta^{2}+\omega_{0}^{2}-\Omega^{2}=2\beta^{2}$), the bracket reduces
to
\begin{equation}
  \omega_{0}^{2}+\beta^{2}\cos2\theta_{\!s}+\beta\Omega\sin2\theta_{\!s}.
  \label{eq:bracket}
\end{equation}
Writing $\beta^{2}\cos2\theta_{\!s}+\beta\Omega\sin2\theta_{\!s}
=\beta\omega_{0}\cos(2\theta_{\!s}-\alpha)$ with
$\alpha=\arctan(\Omega/\beta)$, the \emph{exact} energy is
\begin{equation}
  E(t)=\tfrac{1}{2}MD^{2}e^{-2\beta t}
       \bigl[\omega_{0}^{2}+\beta\omega_{0}\cos(2\Omega t+2\varphi-\alpha)\bigr].
  \label{eq:energy_exact}
\end{equation}

\subsubsection{Weak-Damping Approximation}

The dimensionless ratio $\beta/\omega_{0}=C/(2\sqrt{KM})$ measures how
weakly the system is damped: when $\beta\ll\omega_{0}$ (equivalently
$C^{2}\ll4KM$), this ratio is much less than~1, making it the small
parameter that controls the approximation.
In this regime the oscillating correction in \eqref{eq:energy_exact}
is small and averages to zero over each half-period $T/2=\pi/\Omega$.
Dropping it:
\begin{equation}
  E(t)\approx\tfrac{1}{2}MD^{2}\omega_{0}^{2}e^{-2\beta t}
            =E_{0}e^{-2\beta t},
  \label{eq:energy_approx}
\end{equation}
where $E_{0}=\tfrac{1}{2}KD^{2}$ (using $M\omega_{0}^{2}=K$).
In terms of the original damping constant:
\begin{equation}
  E(t)\approx E_{0}e^{-(C/M)t}.
  \label{eq:energy_C}
\end{equation}
The energy decays at rate $C/M$, twice the amplitude decay rate
$\beta=C/2M$, consistent with the general result for quadratic systems.

\subsubsection{Energy Decay and Quality Factor}
\label{sec:edecay}

The damping ratio is defined as
\begin{equation}
  \xi=\frac{C}{2m\omega_{0}},\qquad \frac{C}{m}=2\xi\omega_{0},
  \label{eq:xi}
\end{equation}
(using $\xi$ rather than $\zeta$ to avoid a symbol clash with the
$\zeta$-plane variable central to this paper).
Equation~\eqref{eq:energy_C} then reads
\begin{equation}
  E(t)=E_{0}e^{-2\xi\omega_{0}t}.
  \label{eq:energy_xi}
\end{equation}
Differentiating:
\begin{equation}
  \frac{dE}{dt}=-2\xi\omega_{0}E,
  \label{eq:dEdt}
\end{equation}
so the energy decay rate is proportional to both $\xi$ and $\omega_{0}$
and decreases continuously over time.

The quality factor is
\begin{equation}
  Q=\frac{1}{2\xi}=\frac{m\omega_{0}}{C},
  \label{eq:Q_def}
\end{equation}
so that \eqref{eq:energy_xi} becomes
\begin{equation}
  E(t)=E_{0}e^{-\omega_{0}t/Q}.
  \label{eq:energy_Q}
\end{equation}
Larger $Q$ means slower energy loss (weak damping); smaller $Q$ means
faster dissipation.

\subsection{Visualisation on the $uv$-Plane}

The circle radius is $R=\omega_{0}t$.
Since $u=\beta t$, we have $R=(\omega_{0}/\beta)\,u=2Qu$, giving
\begin{equation}
  E(R)=E_{0}e^{-R/Q}.
  \label{eq:E_R}
\end{equation}
As the circle in the $uv$-plane expands, the energy decays exponentially
with characteristic length $Q$ in the radial direction.
Per radian of total phase $R$, the oscillator loses a fraction $1/Q$
of its energy---exactly the standard differential definition of $Q$:
\begin{equation}
  Q=2\pi\times\frac{\text{energy stored}}{\text{energy lost per cycle}}.
  \label{eq:Q_std}
\end{equation}
Drawing circles at $R=Q,2Q,3Q,\ldots$ in the $uv$-plane, the energy
at those circles is $E_{0}/e,\,E_{0}/e^{2},\,E_{0}/e^{3},\ldots$


\section{Conclusion}

This work has developed a comprehensive geometric framework for analyzing the underdamped harmonic oscillator by mapping the dynamics onto a logarithmic spiral in the complex plane. The principal results and their implications are as follows.

\textbf{Geometric reformulation of extremum condition.} The displacement extrema of the underdamped oscillator, traditionally found by solving a transcendental equation, become transparent in the \(\zeta\)-plane: they correspond precisely to the instants when the spiral \(\zeta(t)=e^{-i\varphi}we^{-w}\) crosses the imaginary axis. This reformulation yields the explicit time sequence \(t_n = (\theta - \varphi - \pi/2 + n\pi)/\Omega\), where \(\theta = \arctan(\Omega/\beta)\) is the fixed ray angle in the \((u,v)\)-plane. The separation between successive extrema, \(\Delta t = \pi/\Omega\), recovers the damped half-period as expected.

\textbf{Lambert \(W\) function and threshold crossings.} The times at which the spiral radius reaches a specified value \(A\) (e.g., a noise threshold in experimental data) are given by \(t = -\beta^{-1} W_k(-\beta A/\omega_0)\). The two real branches \(k=0\) and \(k=-1\) naturally account for the outward (expanding) and inward (decaying) phases of the motion, respectively. This provides a closed-form analytical solution to a problem that otherwise requires numerical root-finding.

\textbf{Unified geometric interpretation of \(Q\).} The quality factor \(Q = \omega_0/(2\beta)\) is directly encoded in the ray angle \(\theta\) via \(Q = \frac{1}{2}\sec\theta\). As \(\theta\) ranges from \(0\) (critically damped) to \(\pi/2\) (undamped limit), \(Q\) increases from \(1/2\) to infinity. This geometric representation makes the transition from aperiodic to oscillatory behaviour visually intuitive: the spiral collapses to a line segment when \(\theta=0\) and winds infinitely many times as \(\theta\to\pi/2\).

\textbf{Spiral properties as physical diagnostics.} The spiral's winding number \(N_\varepsilon \approx (Q/\pi)\ln(2Q/\varepsilon)\) for large \(Q\) provides a robust method for estimating \(Q\) from experimental data by simply counting visible oscillations above a noise threshold. Unlike the logarithmic decrement, which requires precise amplitude ratios of successive peaks, turn counting remains reliable even when successive amplitudes differ by only a tiny fraction. The enclosed area \(A = \omega_0^2\Omega/(8\beta^3) \approx Q^3\) in the lightly damped limit offers another geometric invariant that scales sensitively with damping.

\textbf{Energy decay in the \(uv\)-plane.} The mechanical energy \(E(t) = \frac{1}{2}MD^2 e^{-2\beta t}[\omega_0^2 + \beta\omega_0\cos(2\Omega t + 2\varphi - \alpha)]\) simplifies in the weak-damping approximation to \(E(t) \approx E_0 e^{-\omega_0 t/Q}\). On the expanding circle of radius \(R = \omega_0 t\) in the \((u,v)\)-plane, the energy decays as \(E(R) = E_0 e^{-R/Q}\), so each increment of \(R\) by \(Q\) reduces the energy by a factor of \(e\). This directly visualises the standard definition of \(Q\) as \(2\pi\) times stored energy divided by energy lost per cycle.

\textbf{Unification across damping regimes.} The \(\zeta\)-plane spiral naturally encompasses all three damping regimes: underdamped (\(Q>1/2\)) gives a genuine spiral winding about the origin; critically damped (\(Q=1/2\)) collapses to a line segment; overdamped (\(Q<1/2\)) remains a line segment but with different parametrisation. This unified representation makes the transition between regimes continuous and geometrically transparent.

\textbf{Pedagogical and practical value.} The framework presented here offers an accessible entry point to several advanced mathematical concepts---complex variables, the Lambert \(W\) function, winding numbers, and conformal mapping---within the familiar context of the harmonic oscillator. For practitioners, the turn-counting method provides a noise-robust alternative to traditional techniques for estimating \(Q\) in high-\(Q\) systems, while the closed-form Lambert \(W\) expressions enable efficient computation of threshold crossing times without numerical iteration.

\textbf{Future directions.} The geometric approach developed here may extend naturally to driven damped oscillators (where the spiral would acquire an additional time-dependent forcing term), to nonlinear systems\cite{calasananalytical} (where the spiral would distort into more complicated curves), and to coupled oscillators (where higher-dimensional analogues would involve multiple interacting spirals). The connection between the Lambert \(W\) function and the spiral radius threshold suggests potential applications in signal processing, control theory, and system identification where rapid estimation of damping parameters from noisy data is required.

In summary, the transformation \(t \mapsto \zeta = e^{-i\varphi}we^{-w}\) converts the underdamped harmonic oscillator into a logarithmic spiral whose geometry directly encodes all dynamical quantities of interest. This unification of classical mechanics, complex analysis, and special functions provides both a powerful analytical tool and an intuitive geometric picture of damped oscillations.
%
\bibliographystyle{unsrt}
\bibliography{references}
\balance
\end{document}